\documentclass[onecolumn,notitlepage,pre,aps,floatfix,superscriptaddress,longbibliography]{revtex4-1}
\pdfoutput=1 
\usepackage{amsmath,amssymb,eucal,graphicx,float,epstopdf,xparse}
\usepackage[colorlinks=true, urlcolor=blue, anchorcolor=blue, citecolor=blue,filecolor=blue,linkcolor=blue,menucolor=blue]{hyperref}

\usepackage{url}
\usepackage{graphicx}
\usepackage{dcolumn}
\usepackage{amssymb}
\usepackage{bm}
\usepackage{color}
\usepackage{enumitem}
\usepackage{mathtools}
\usepackage{comment}
\usepackage{subfigure}
\usepackage{float}

\begin{document}
\title{Epidemic Forecast Follies}
\author{P. L. Krapivsky}
\affiliation{Department of Physics, Boston University, Boston, MA 02215, USA}
\affiliation{Santa Fe Institute, 1399 Hyde Park Road, Santa Fe, NM 87501, USA}
\author{S. Redner}
\affiliation{Santa Fe Institute, 1399 Hyde Park Road, Santa Fe, NM 87501, USA}

 \begin{abstract}
   We introduce a simple multiplicative model to describe the temporal
   behavior and the ultimate outcome of an epidemic.  Our model
   accounts, in a minimalist way, for the competing influences of
   imposing public-health restrictions when the epidemic is severe,
   and relaxing restrictions when the epidemic is waning.  Our primary
   results are that different instances of an epidemic with identical
   starting points have disparate outcomes and each epidemic temporal
   history is strongly fluctuating.
\end{abstract}

\maketitle

\section{Background}

Now that the most severe (we hope) manifestations of the Covid-19
epidemic have passed, one can't help but realize that many of the
early forecasts of the Covid-19 epidemic toll were wildly inaccurate
and inconsistent with each other.  Moreover, individual forecasts
could change dramatically over a period of few days.  For the USA, in
particular, the earliest estimates for the Covid-19 epidemic death
toll ranged from tens of thousands to many millions, with the current
death toll (as of September 2023) reported to be 1.175 million out of
a total of 108.5 million cases (all data taken from~\cite{Covid}).
Perhaps even more striking are the huge fluctuations and the
dramatically different time courses in the daily death rate in
different countries.

\begin{figure}[ht]
  \centerline{\subfigure[]{\includegraphics[width=0.45\textwidth]{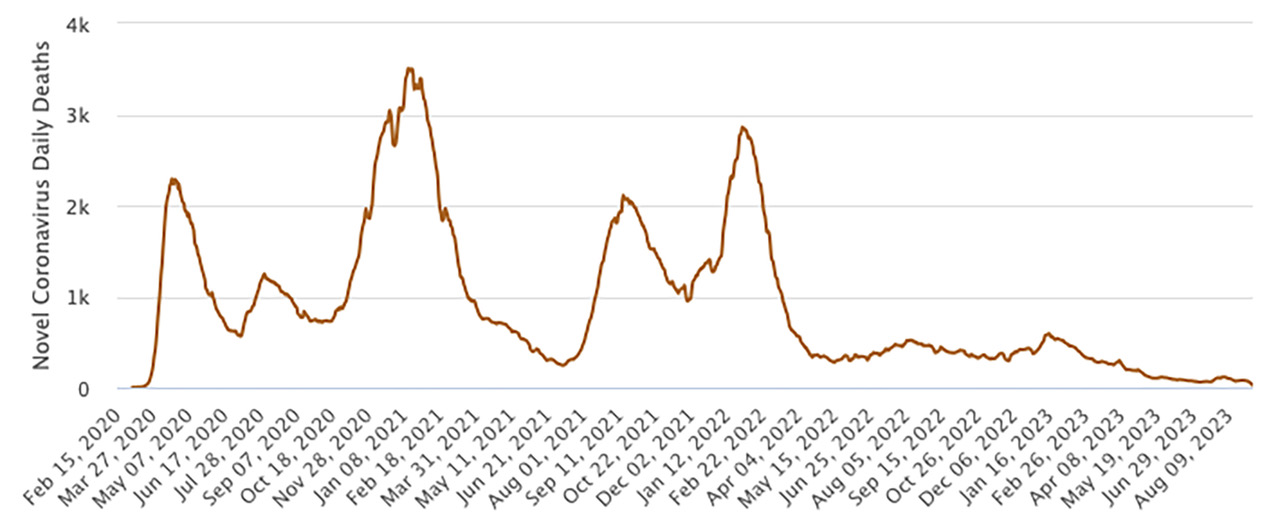}}\qquad
    \subfigure[]{\includegraphics[width=0.45\textwidth]{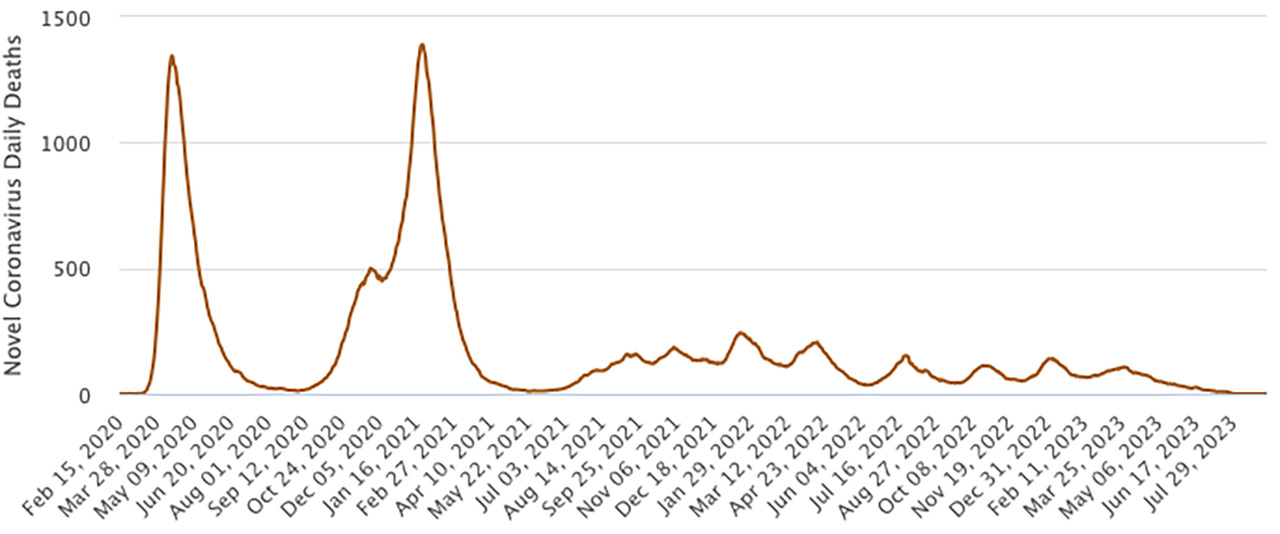}}}
  \centerline{\subfigure[]{\includegraphics[width=0.45\textwidth]{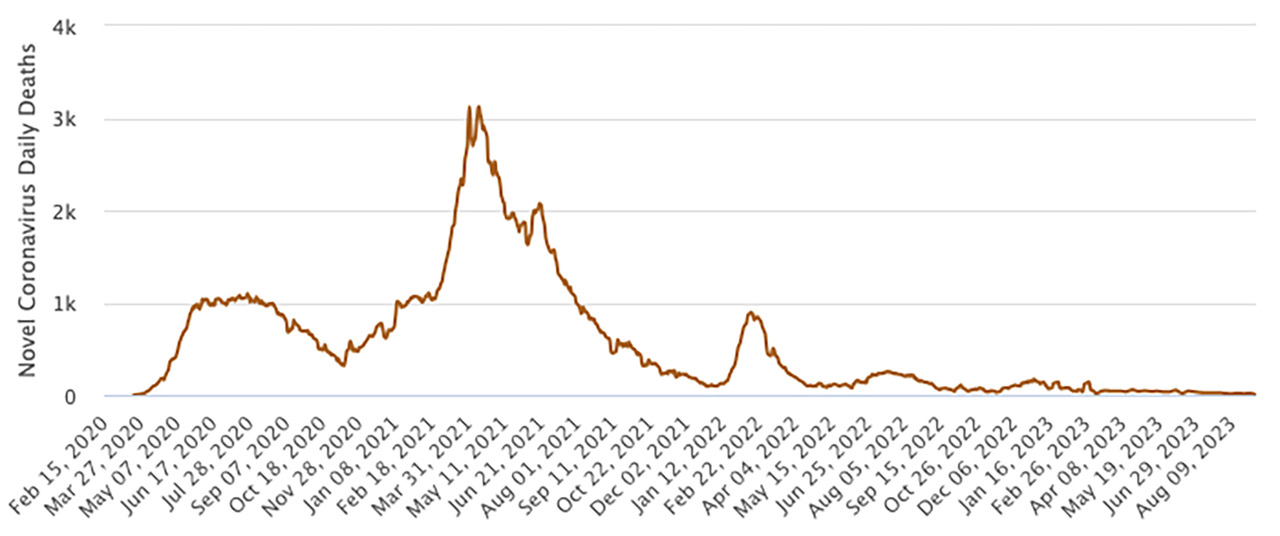}}\qquad
     \subfigure[]{\includegraphics[width=0.45\textwidth]{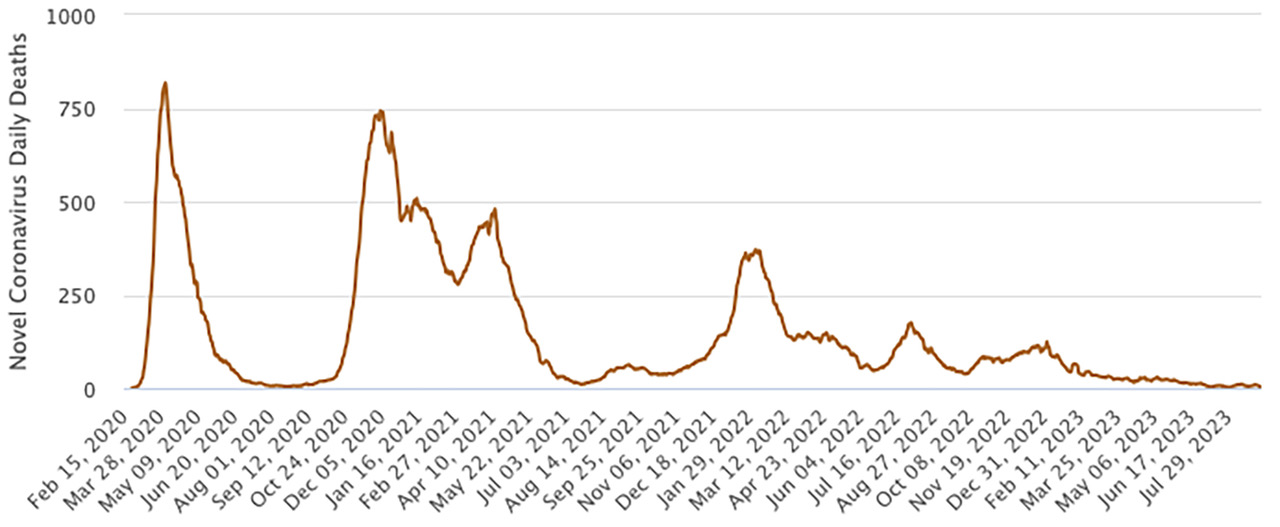}}}
\centerline{\subfigure[]{\includegraphics[width=0.45\textwidth]{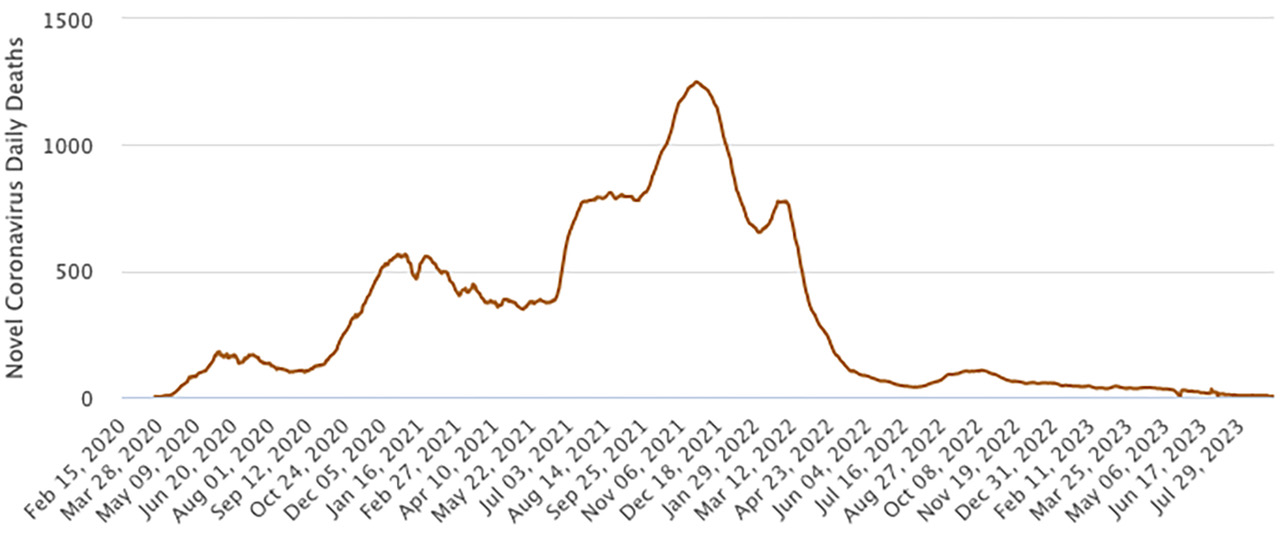}}\qquad
   \subfigure[]{\includegraphics[width=0.45\textwidth]{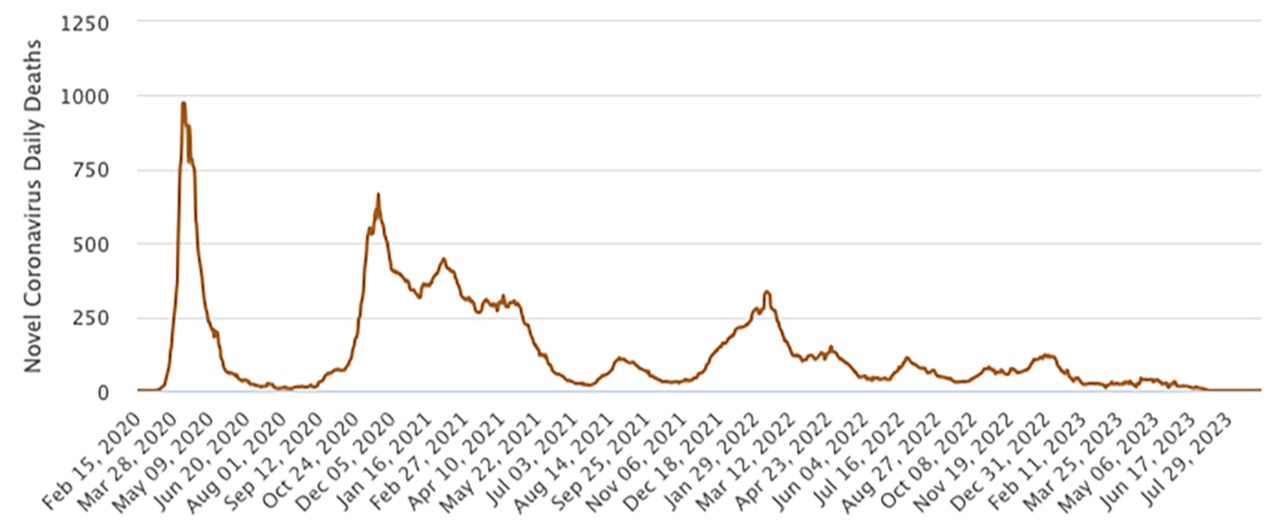}}}
\caption{\small The reported time dependences of the daily
  Covid death rates (7-day moving average) for the (a) USA, (b) UK,
  (c) Brazil, (d) Italy, (e) Russia,
  and (f) France.  These data cover the period from Feb.\ 15 2020 until July 29, 2023
  and are all taken from Ref.~\cite{Covid}.
  \label{fig:death}}
\end{figure}

To illustrate these statements, Fig.~\ref{fig:death} plots the
reported daily death rates for the six countries in the world with
populations greater than 60 million and with the largest total death
rates.  They are: USA (3.507 deaths/1000), UK (3.339/1000), Brazil
(3.275/1000), Italy (3.174/1000), Russia (2.743/1000), and France
(2.556/1000).  For reference, the country with the largest reported
total death rate is Peru (6.582/1000), while the world average is
(0.887/1000).  For many reasons, the accuracy of the data may vary
widely from country to country so that some of the numbers reported in
Ref.~\cite{Covid}, such as the suspicious smoothness of the data for
Russia, should be interpreted with caution.

One of the many confounding features of Covid-19 is asymptomatic
transmission, in which the epidemic may be unknowingly spread by
individuals who did not know that they were contagious.  Partly
because of this feature, a wide variety of increasingly sophisticated
multi-compartment models were developed that build on the classic SIR
and SIS models of epidemic spread.  These models typically attempted
to faithfully account for subpopulations in various stages of the
disease and recovery, as well as the transitions between these stages.
Models of this type gave rise to complex dynamical behaviors that
could sometimes mirror reality in a specific setting or over a limited
time range.  However, embellishments of SIR and SIS-type models still
seem to be incomplete because of the difficulty in simultaneously
accounting for both the disease dynamics and its interaction with
social forces.

The discrepancy between the observed wildly varying features of
Covid-19 and supposedly deterministic outcomes of SIR and SIS models
is especially striking.  In fact the determinism of the SIR and SIS
models is actually illusory.  The SIR model, for example, is an
inherently stochastic process~\cite{Bailey-SIR,Bailey} that is
characterized by the reproductive number $R_0$.  This quantity is
defined as the average number of individuals to whom a single infected
individual transmits the infection before this single individual
recovers.  In the supercritical regime, $R_0>1$, it is possible that
the outbreak may quickly die out.  This happy event occurs with
probability $R_0^{-1}$ if one individual was initially infected.
Otherwise, the infection quickly spreads, and the behavior becomes
effectively deterministic.  Namely a finite fraction $c=c(R_0)$
individuals catch the disease, with $c$ implicitly determined by the
criterion~\cite{Hethcote}
\begin{equation}
c+ e^{-cR_0}=1
\end{equation}

Conversely, if $R_0<1$, the outbreak quickly dies out, so while the
subcritical SIR process is still manifestly stochastic, it is not a
threat to the population at large.  The interesting and the most
strongly stochastic behavior emerges in critical SIR and SIS
models~\cite{Ridler-Rowe,Grassberger,Martin,BK04-SIR,Kessler07,Kessler08,Gordillo08,Greenwood2009,Hofstad,Antal-SIR,BK12-SIR,K21-SIR}.
For the SIR mode, in particular, the distribution of the number of
infected individuals has a power-law tail.  For a finite population of
size $N$ the critical SIR model does not lead to a pandemic, because
the average number of individuals who contract the disease scales as
$N^{1/3}$.

Here we argue that significant forecasting uncertainties are an
integral feature of processes caused by the interplay between the
dynamics of the disease transmission and the social forces that arise
in response to the epidemic.  Each attribute alone typically leads to
either exponential growth (due to disease transmission at early times)
or to exponential decay (due to effective mitigation strategies).
Within our model, the competition between these two exponential
processes leads to a dynamics that is extremely sensitive to seemingly
minor details.  The basic mechanism in our modeling is that the
reproductive number $R_0$ can sometimes decrease, due to the
imposition of public-health measures, such as social distancing,
vaccinations, etc., and sometimes increase, because of the relaxation
of these measures.  Focusing only on the dynamics of the reproductive
number serves as a useful proxy for the myriad of influences that
control the true epidemic dynamics.  Within this framework, we will
determine the duration of an epidemic, the time dependence of the
number of infected individuals, and the total number of individuals
infected when an epidemic finally ends.  All three quantities exhibit
huge fluctuations that are reminiscent of the actual data.

\section{Systematic Mitigation}

As a preliminary, we first investigate what we term as the systematic
mitigation strategy. Here increasingly stringent controls are imposed
as soon as an outbreak is detected until the reproductive number $R_0$
is reduced to below 1.  Once $R_0<1$, progressively fewer individuals
are infected after each incubation period, so that the epidemic soon
disappears.  The condition $R_0=1$ defines the end of the epidemic.
Because society is a complicated, with many competing social forces in
play, we posit that it is not possible to reduce $R_0$
instantaneously, but rather, the reduction happens gradually.  We
therefore assume that after each successive incubation period $R_0$ is
decreased by a random number $r$ whose average value
$\langle r\rangle$ is less than 1.  While additional individuals will
become infected after $R_0$ has been reduced to less than 1, their
number decays exponentially with time and constitute a negligible
contribution to the total number of infections.

\begin{figure}[ht]
  \centerline{
   \includegraphics[width=0.425\textwidth]{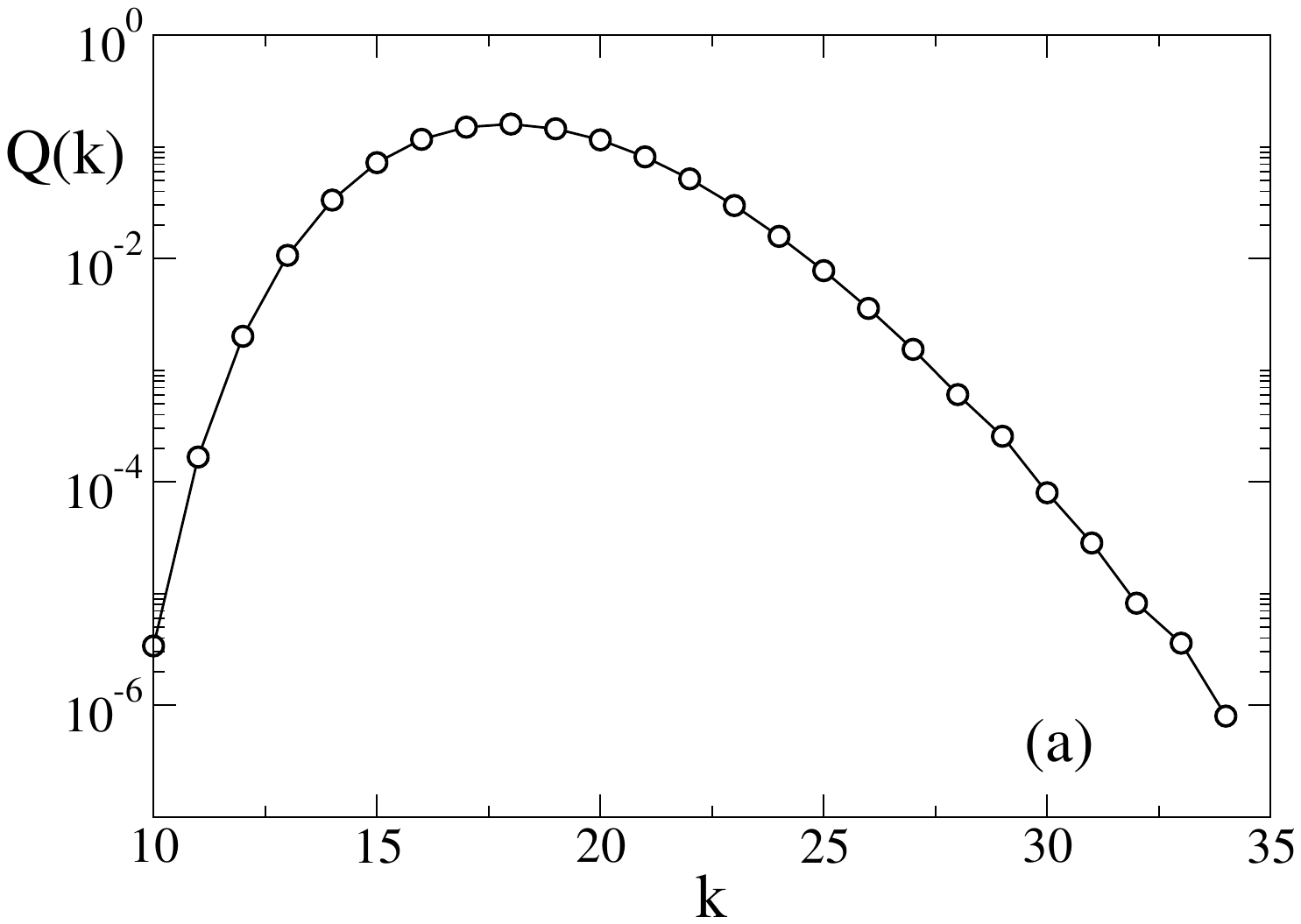}\qquad
  \includegraphics[width=0.425\textwidth]{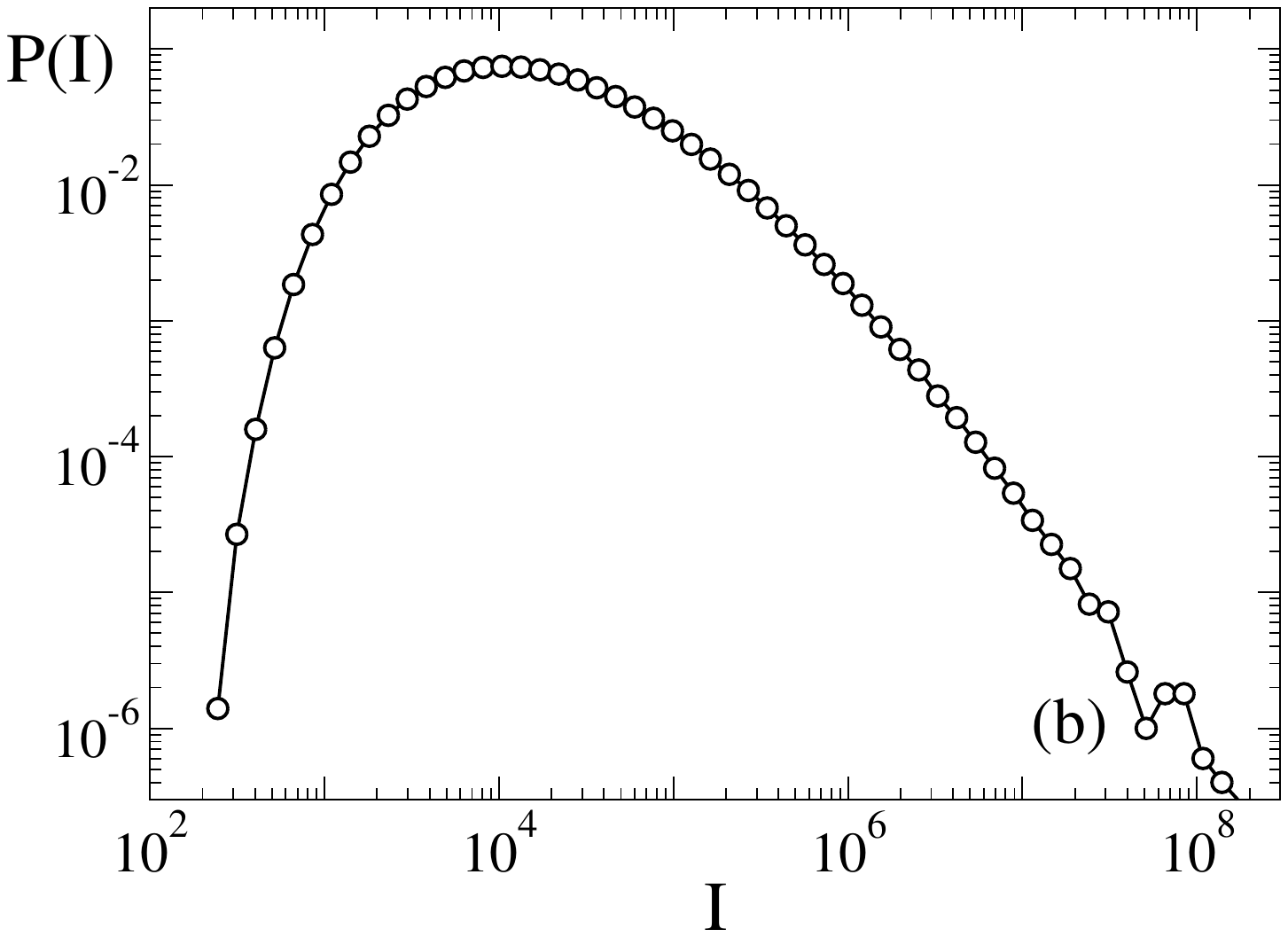}
  }
  \caption{Systematic mitigation: (a) The probability $Q(k)$ that the
    epidemic lasts $k$ periods.  (b) The probability $P(I)$ that the
    epidemic has ultimately infected $I$ people (under the assumption
    that the initial epidemic size is one person).}
\label{fig:tn}  
\end{figure}

Define $R_k$ as the reproductive number on the $k^{\rm th}$ period.  Then
$R_k$ is given by
\begin{align}
  R_k= r_k \,R_{k-1}= r_k\, r_{k-1}\ldots r_2\,r_1\, R_0\,,
\end{align}
where $r_k$ is the value of the random variable $r$ in the
$k^{\rm th}$ period. The typical number of periods $k$ until $R_0$
reaches 1 is determined by $R_0 \,\langle r\rangle^k = 1$. In what
follows, we assume that when the epidemic is first detected, the
reproductive number $R_0=2.5$, and we take $\langle r\rangle = 0.95$
for illustration. With these conventions,
\begin{equation*}
k=\frac{\ln (1/R_0)}{\ln \langle r\rangle}=\frac{\ln (1/2.5)}{\ln(0.95)}\approx 17.86
\end{equation*}
Thus the epidemic typically disappears after 18 periods.  However,
because of the inherent randomness in the mitigation, with $R_0$
sometimes decreasing by less than 0.95 and sometimes by more than
0.95, the true epidemic dynamics can be very different, as illustrated
in Fig.~\ref{fig:tn}.

We simulate the systematic mitigation strategy by starting with a
single infected individual and reproductive number $R_0=2.5$.  We then
choose a set of random numbers $r_1, r_2, r_3,\ldots$, each of which
are uniformly distributed between 0.9 and 1, so that
$\langle r\rangle =0.95$.  We first measure how long it takes until
$R_k$, the reproductive number in the $k^{\rm th}$ period, is reduced
to 1, which signals the end of the epidemic.  We perform this same
measurement for $5\times 10^6$ different choices of the set of random
numbers $r_1, r_2,\ldots, r_k$.  As shown in Fig.~\ref{fig:tn}(a), the
probability $Q(k)$ that the epidemic is extinguished in $k$ periods is
peaked at roughly 18 periods, consistent with the naive estimate
above.  If one is lucky, that is, if most of the reduction factors
$r_i$ are close to 0.9, the epidemic can be extinguished in as little
as 11 periods.  If one is unlucky (many of the $r_i$ are close to 1),
the epidemic can last more than 30 periods.

While the distribution of epidemic durations is fairly narrow, the
size of an epidemic, namely, the total number $I$ of people who were
infected during the course of an epidemic,
\begin{align}
\label{I:sum}
  I=R_0 +R_1+R_2+\ldots+ R_k\,,
\end{align}
can vary by several orders of magnitude. It is important to point out
that the number of newly infected people is based on the assumption
that this number is small compared to the total population size, so
that the growth in the number of new infections is truly exponential.
As shown in Fig.~\ref{fig:tn}(b), while the most probable epidemic
size is $\approx 10^4$ (again starting with a single infected
individual), there is a non-vanishing probability that the outbreak
size can be as small as a few thousand or greater than $10^7$.  This
large disparity in outbreak sizes illustrates how small changes in the
way that the epidemic is mitigated can lead to huge changes in the
outbreak size.

More dramatically, suppose that the mitigation strategy is slightly
less effective and that the reproductive number is reduced at each
period by a uniform random variable that lies between $[0.95,1]$
rather than between $[0.9,1]$.  Now the epidemic can last between 22
and 55 periods, with a most probable duration of 36 periods.  However,
the epidemic size ranges between roughly $10^5$ and $10^{12}$, with a
most probable size of roughly $7\times 10^7$.  The upper value is much
larger than the world population and the finiteness of the population
would now provide the upper bound.  Although this second epidemic
lasts twice as long as the first one, it typically infects 7,000 times
more people!  We emphasize that the stochastic nature of the random
variables $r_j$ plays a decisive role. Very different behaviors emerge
in the deterministic case \cite{Ginestra}.

We also mention that the systematic mitigation strategy is
analytically tractable because of a close relation between the
epidemic size in \eqref{I:sum} and Kesten variables~\cite{Kesten73},
which appear in probability theory, one-dimensional disordered
systems, and various other subjects.  We explain this connection in
Appendix~\ref{ap:Kesten} and also several analytical results that
qualitatively agree with our numerical observations.  As one example,
we show that the slightly faster than exponential decay of $P(I)$
suggested by Fig.~\ref{fig:tn} may be close to a factorial decay.

\section{Vacillating Mitigation}

During the acute period of the pandemic in 2020--2021, there was
considerable and even vitriolic debate about the efficacy of various
mitigation strategies, or even about the utility of any mitigation.
If the epidemic is severe, as quantified by the reproductive number
$R_k$ in the $k^{\rm th}$ period being substantially greater than 1,
people may be more likely to accept restrictions on their behaviors,
such as isolating, masking, vaccinating, etc., to reduce their risk of
getting sick.  These adaptations will reduce the reproductive number.
If, however, the reproductive number becomes less than 1, then people
will want to relax their vigilance and may also advocate for the
opening of various public venues, such as schools, theaters, stadiums,
etc.  We model this tug-of-war between increased and decreased
restrictions by what we term as the vacillating mitigation strategy.
This perspective of treating the competition between epidemiology and
social behavior was previously treated in more sophisticated
models~\cite{Manrubia,Maslov}.  We emphasize that our model merely a
proxy for the two competing influences of epidemiology and social
behavior.

The two competing steps of the vacillating strategy are the following:
\begin{itemize}
  \itemsep -0.5ex
\item Mitigation: if $R_k>1$, decrease $R_k$ by a factor $r$ that is
  uniformly distributed in $[a,1]$, with $a<1$.

\item Relaxation: if $R_k<1$, change $R_k$ by a factor $s$ that is
  uniformly distributed in $[a,3-2a]$.
\end{itemize}
The first option is the same as in the systematic mitigation strategy.
We construct the second option by requiring that
$\langle s\rangle=1+\frac{1}{2}(1-a)$ and
$\langle r\rangle=1-\frac{1}{2}(1-a)$ are symmetrically located about
1.  That is, the average decrease in $R_k$ in a mitigation step equals
the average increase in $R_k$ in the relaxation step.  This
symmetrical construction seems appropriate to probe the long-term
influence of vacillation on the dynamics.  If the vacillation strategy
was biased towards relaxation, $R_0$ would remain greater than 1 and
the entire planet would be infected.  If this strategy was biased
towards mitigation, the epidemic would be similar to that in
systematic mitigation.  Neither of these cases is interesting from the
viewpoint of probing long-time behaviors.

\begin{figure}[ht]
  \centerline{
   \includegraphics[width=0.425\textwidth]{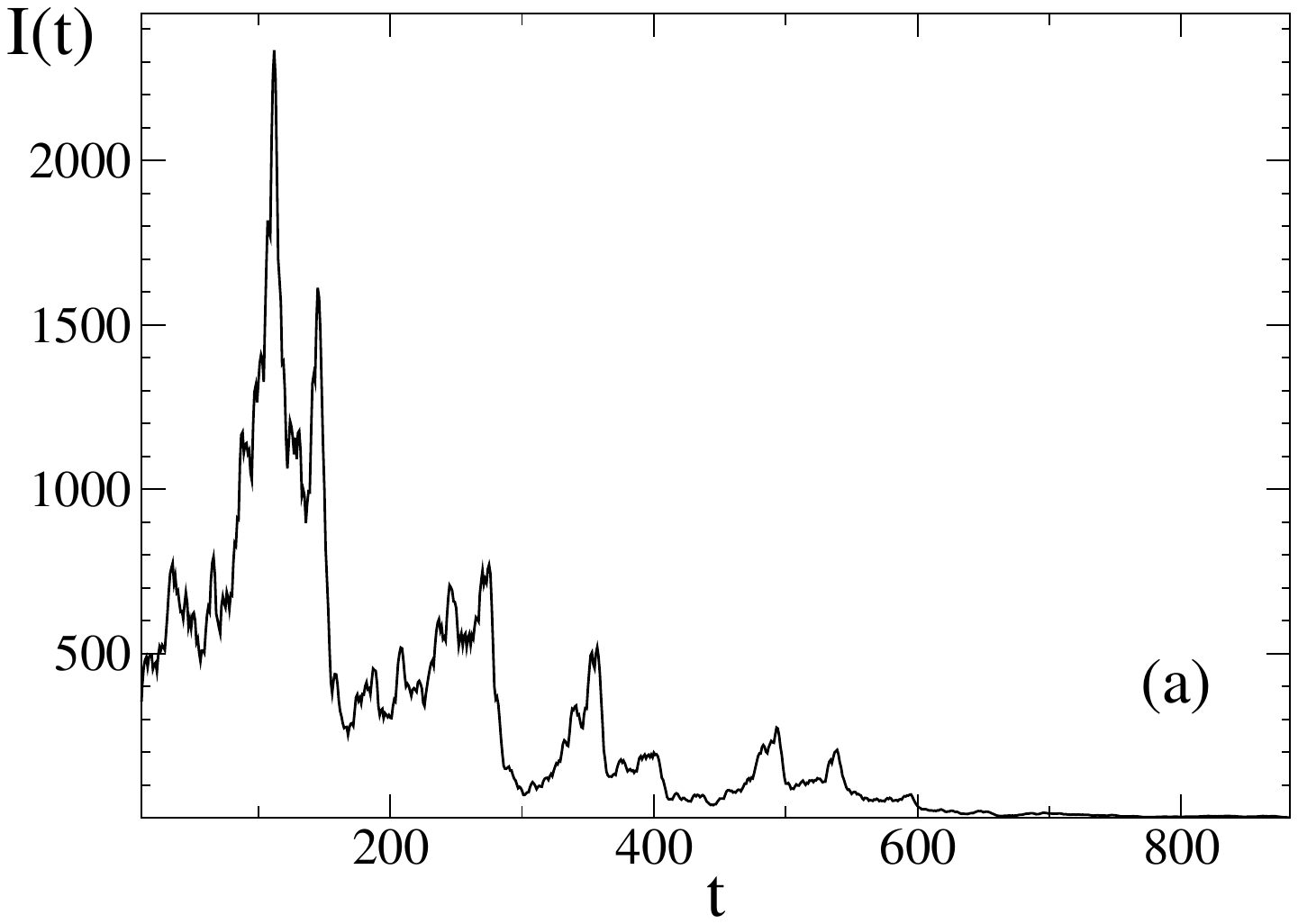}\qquad
    \includegraphics[width=0.425\textwidth]{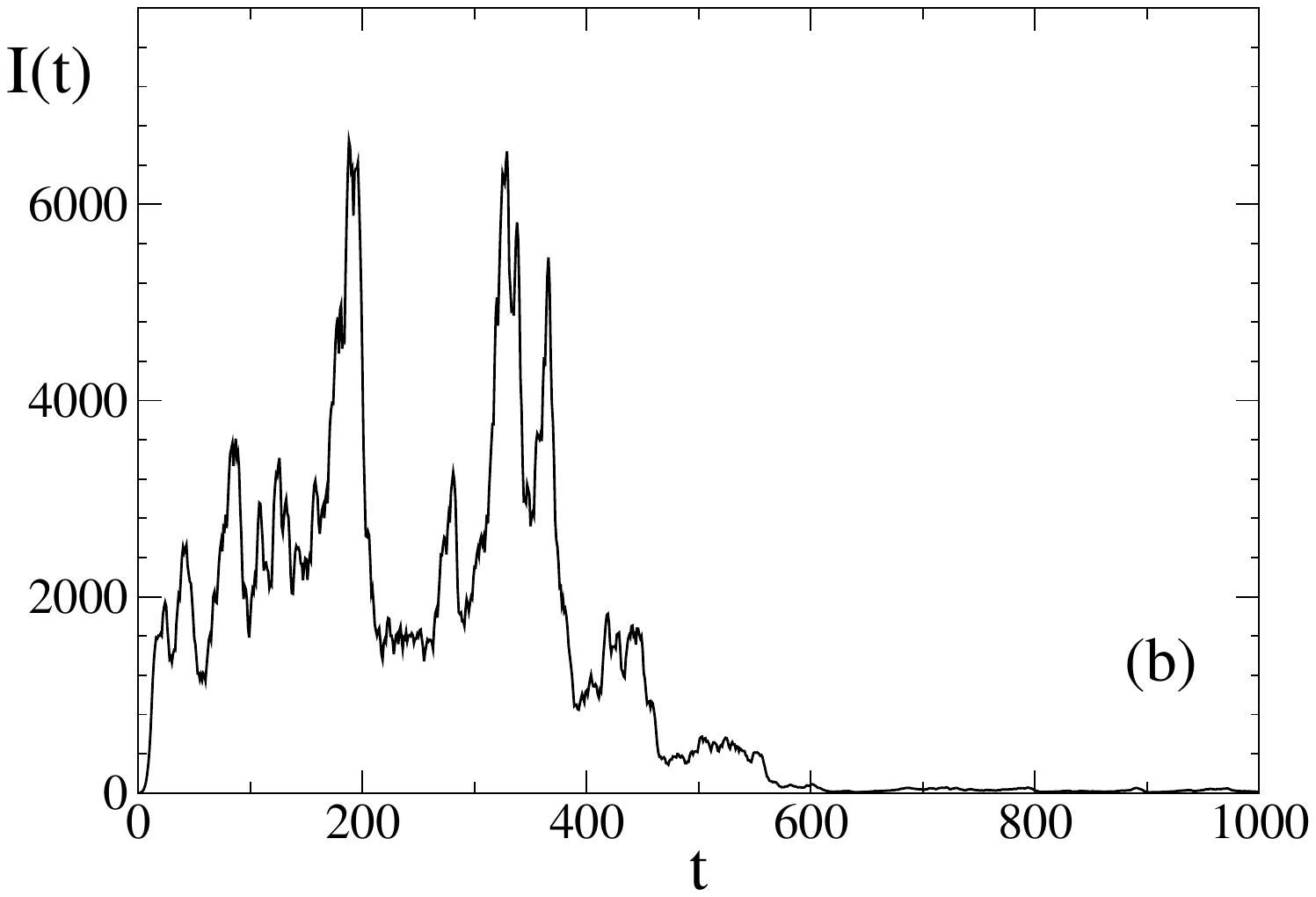}}
  \centerline{ \includegraphics[width=0.425\textwidth]{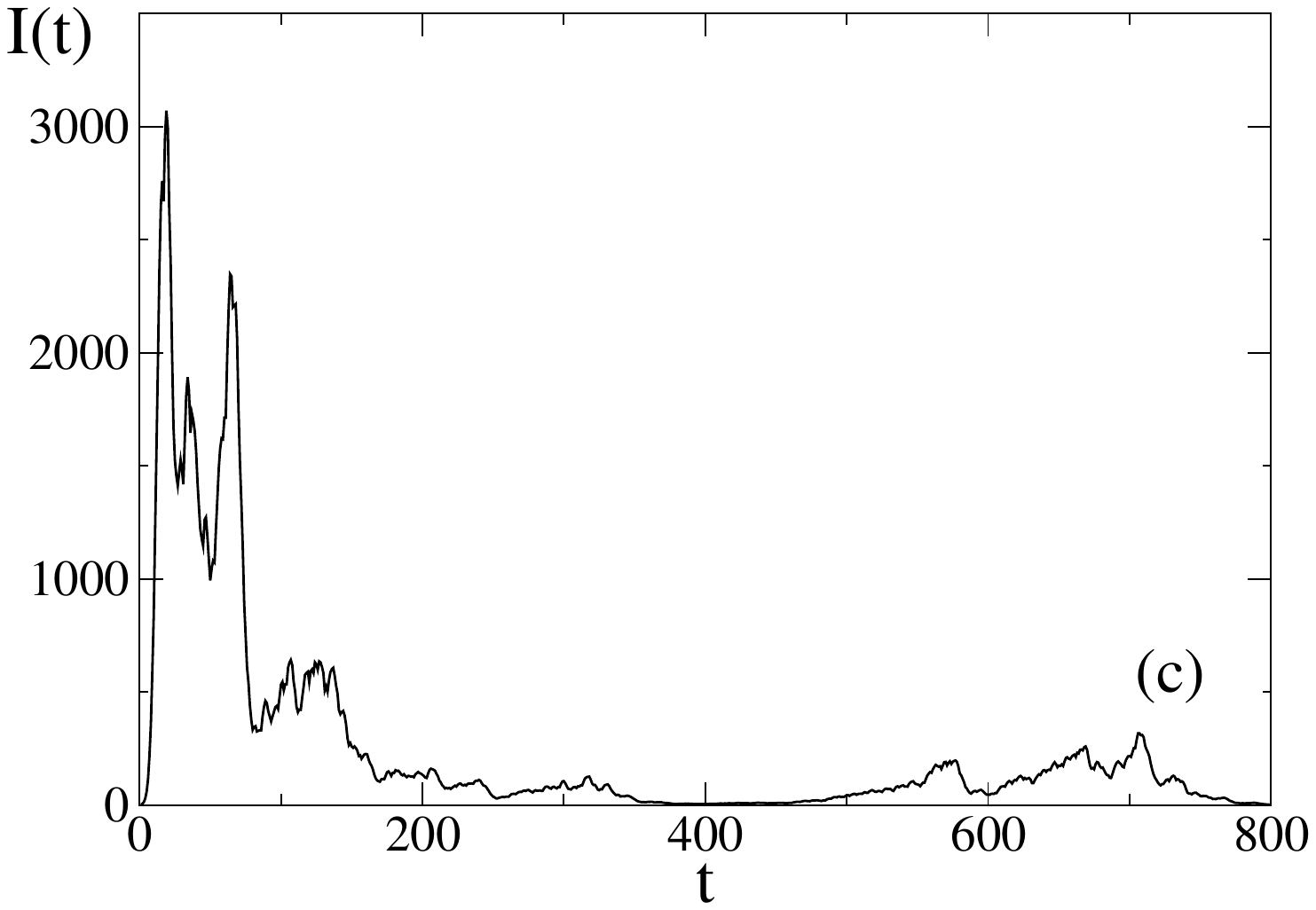}\qquad
  \includegraphics[width=0.425\textwidth]{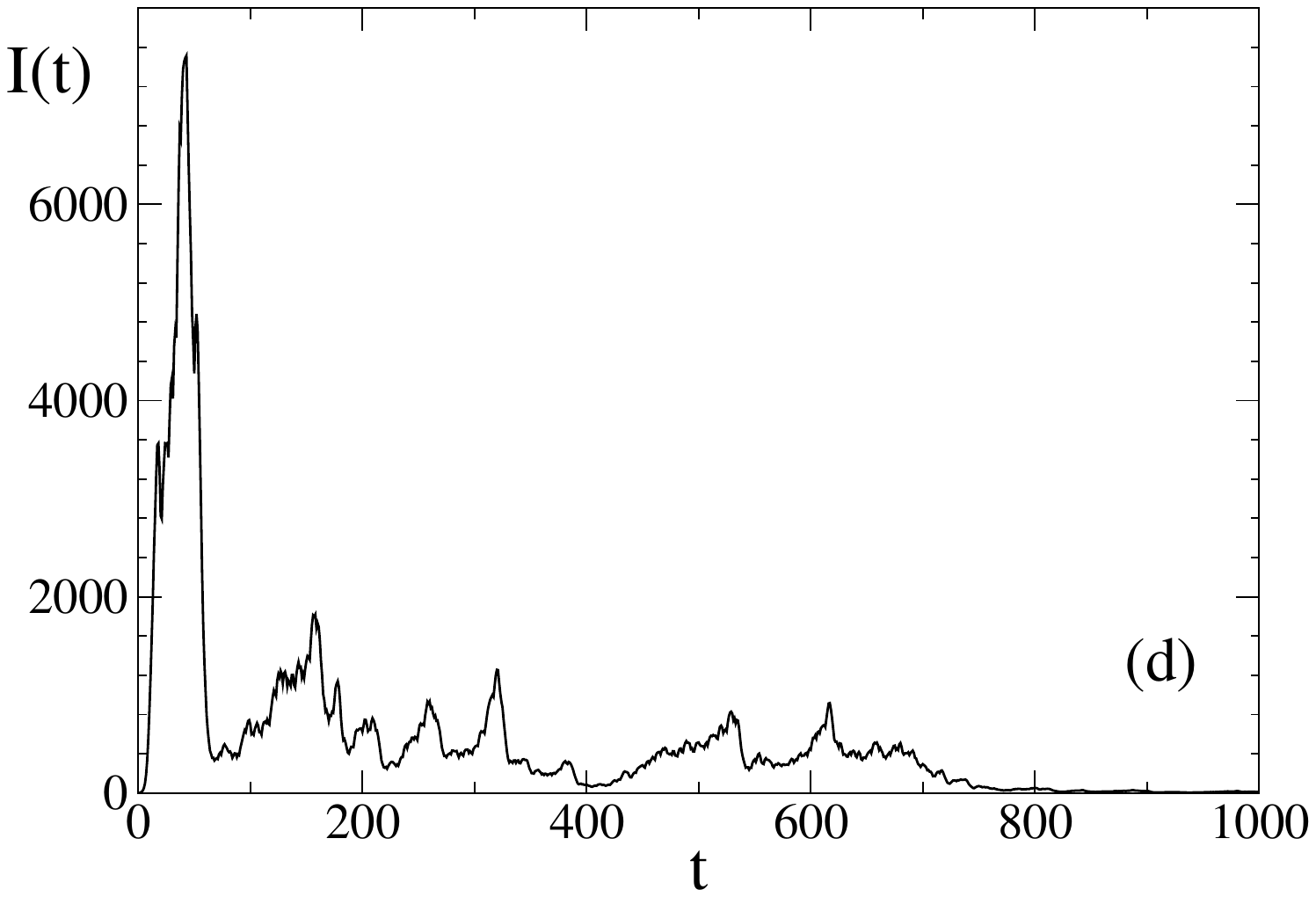}}
\caption{Representative trajectories for the number of people $I(t)$
  infected at time $t$ for the vacillating mitigation strategy when
  starting with $R_0=1$ and a single infected person.}
\label{fig:ntv}  
\end{figure}

In this vacillating strategy, $R_k$ varies between values greater than
1 and values less than 1.  This would lead to an eternal epidemic.  To
avoid this unrealistic outcome, the other important feature of the
relaxation step is that the value of $R_k$ could still decrease during
a relaxation step because $a<1$.  This possibility ensures that
eventually less than one person will be infected in the current
incubation period.  We now define this event as signaling the end of
the epidemic.

Figure~\ref{fig:ntv} shows a few representative trajectories of the
number of people infected $I(t)$ as a function of time (incubation
periods) from the same initial condition of a single infected person
and $R_0=2.5$.  While there are some qualitative differences between
the trajectories of Fig.~\ref{fig:death} and the model outcomes, the
important points that are common to the real data and the simulation
results are the disparities in the individual trajectories and the
strongly fluctuating temporal behavior.

\begin{figure}[ht]
  \centerline{
   \includegraphics[width=0.4\textwidth]{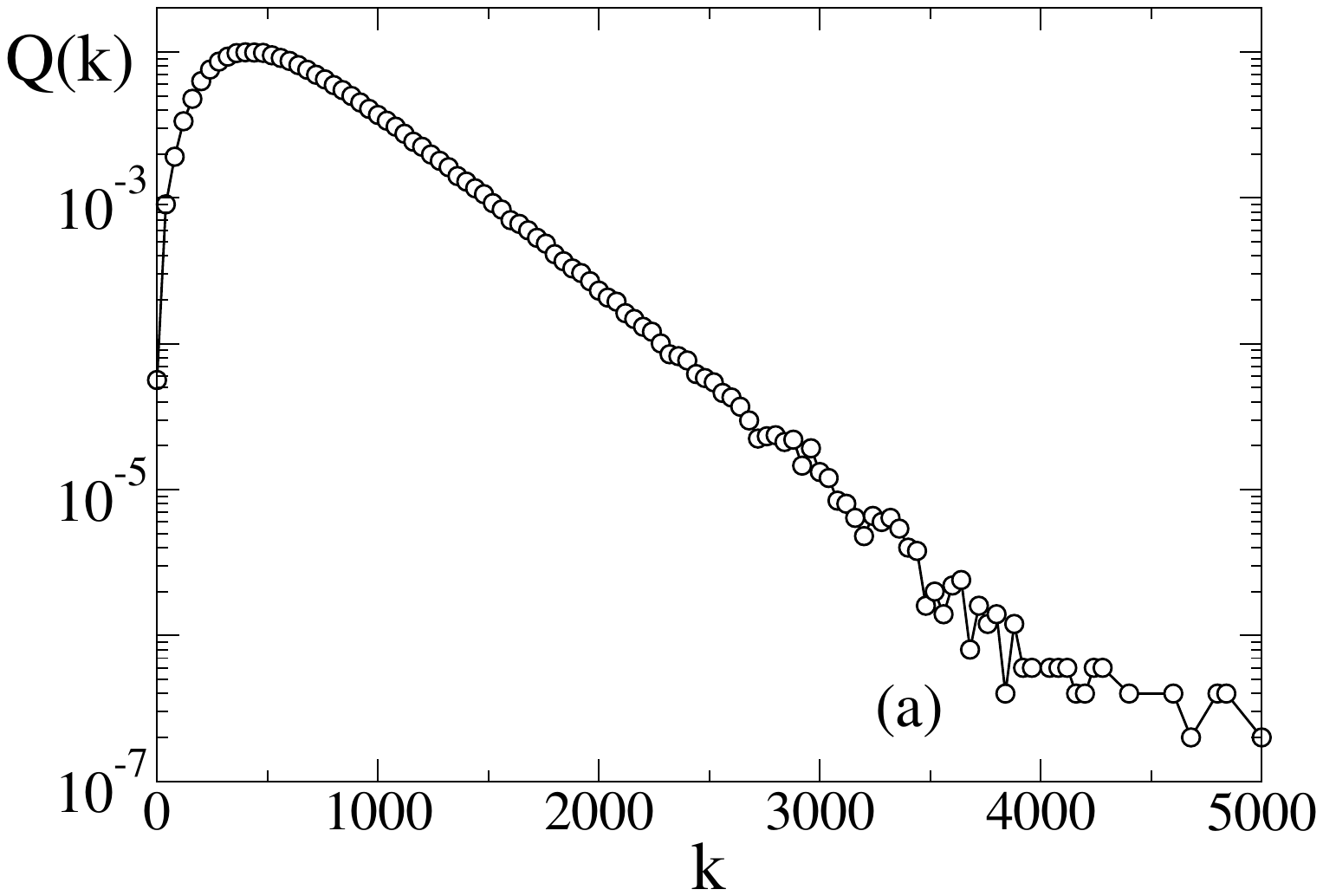}\qquad\qquad
    \includegraphics[width=0.4\textwidth]{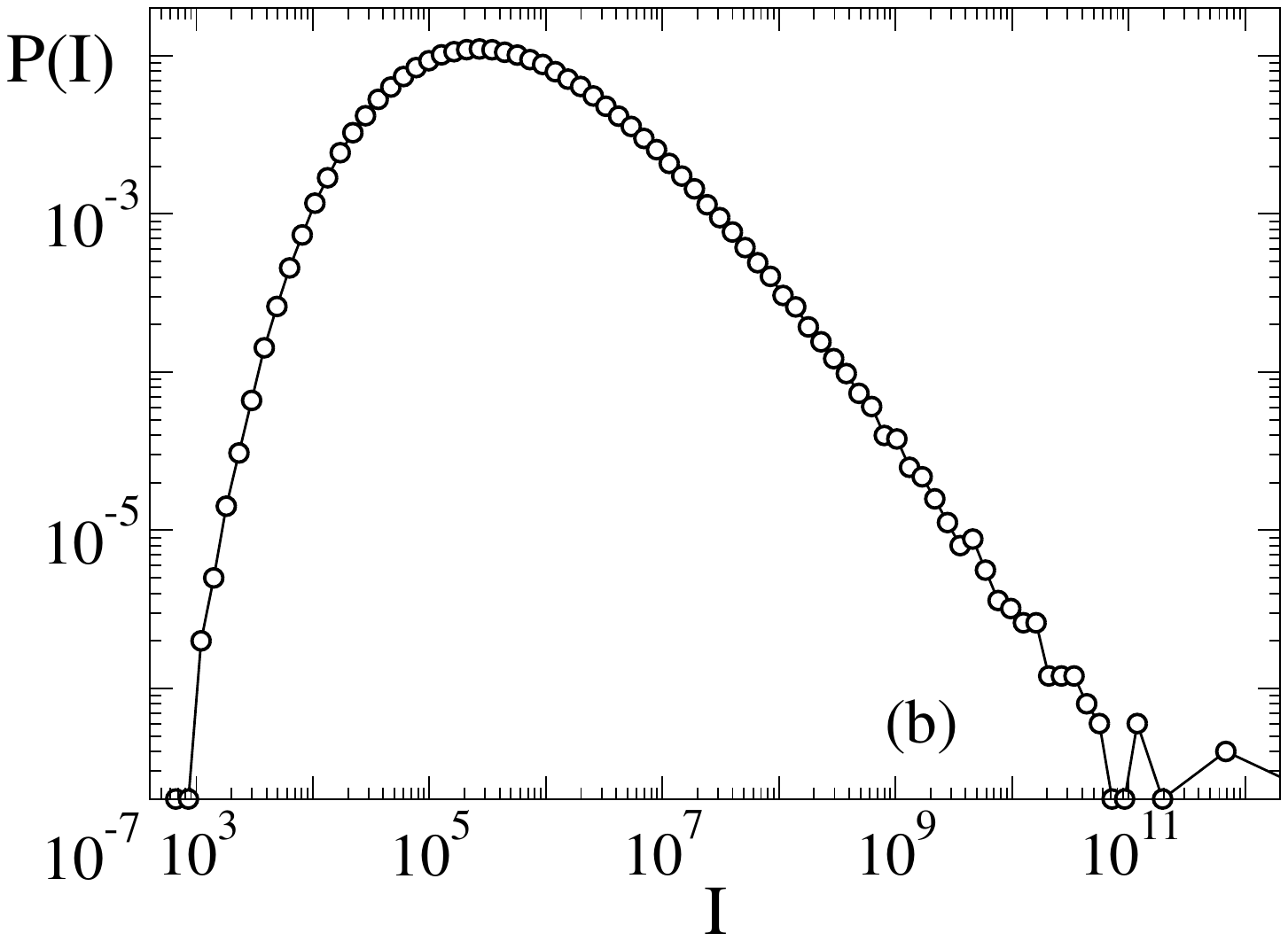}
  }
  \caption{Vacillating mitigation: (a) The probability $Q(k)$ that the
    epidemic lasts $k$ periods.  (b) The probability $P(s)$ that the
    epidemic has ultimately infected $s$ people (under the assumption
    that the initial epidemic size is one person).}
\label{fig:tnv}  
\end{figure}

For the vacillating strategy and for the choice $a=0.9$, the most
likely duration of the epidemic is roughly 400 periods
(Fig.~\ref{fig:tnv}(a)), compared to 18 periods for the systematic
strategy.  The probability that the epidemic lasts much longer than
the most likely value decays exponentially with time.  An even more
dramatic feature of the vacillating strategy is the number of people
that are ultimately infected.  The most probable outcome is that
$3\times 10^5$ people are infected when the epidemic ends
(Fig.~\ref{fig:tnv}(b)).  However, the typical size of the epidemic
can range from $10^4$ to $10^8$.  Compared to the systematic
mitigation strategy with a reduction factor uniformly in the range
$[0.9,1]$, the epidemic now lasts roughly 20 times longer and infects
a factor 30 more individuals.

\section{Concluding Remarks}

This work should not be construed to mean that public-health measures
should be ignored.  Indeed, the extremely rapid development of a
vaccine that is effective against Covid-19 is an outstanding triumph
of modern medical science.  It should also be pointing out that some
of the many forecasting models for Covid-19 were useful during the
early stages of the pandemic.  However, when social influences with
competing viewpoints began to dictate individual and collective policy
decisions, much of the predictive power of forecasting models was
lost.

We also emphasize that our simplistic model has little connection to
the actual epidemiological and social processes that determine the
spread of the epidemic and the changes in individual and collective
behaviors in response to the epidemic.  Nevertheless, our model seems
to capture the tug of war between public-health mandates to control
the spread of the disease and the social forces that often advocate
for a more laissez-faire approach.  Our main message is that there are
huge uncertainties in predicting the time course of an epidemic, its
ultimate duration, and the final outbreak size.  This unpredictability
seems to be intrinsic to the dynamics of epidemics where
epidemiological influences occur in concert with social forces.  In
this setting, forecasting ambiguity is unavoidable.

\bigskip\noindent We thank J. M. Luck for helpful correspondence.
This research was partially supported by NSF grant DMR-1910736.

\appendix
\section{Kesten variables}
\label{ap:Kesten}

We outline an analytical treatment of the systematic mitigation
strategy.  Since the terms in the sum in Eq.~\eqref{I:sum} decay
exponentially in the number of factors in the product, we can replace
the finite sum in \eqref{I:sum} by the infinite sum
\begin{equation}
\label{R:sum}
R = 1 + r_1 + r_1r_2 + r_1 r_2 r_3 +\ldots, \qquad R = I/R_0 \,,
\end{equation}
because it changes the outcome just by a finite number. Random
variables that are defined by \eqref{R:sum} are known Kesten
variables, which have fundamental
implications~\cite{Kesten73,Solomon,Vervaat} and a variety of
applications~\cite{Derrida83,JML85,JML88,Nieuwenhuizen,Poland,PLD,Satya}.

We now show how to probe the probability distribution $P(R)$ using
Kesten variables. The definition of Kesten variables implies that
$P(R)$ satisfies the integral equation
\begin{equation}
\label{Kesten:int}
P(R) = \int dr\,\rho(r)\int dQ\,P(Q)\,\delta[R-1-rQ]\,.
\end{equation}
By performing the Laplace transform
\begin{equation}
\label{Laplace}
\widehat{P}(s)= \int_1^\infty dR\,P(R)\,e^{-sR}\,,
\end{equation}
the Laplace transform of the probability distribution $P(R)$ can be
expressed as
\begin{equation}
\label{Laplace:int}
\widehat{P}(s)= \int_1^\infty dQ\,P(Q)\int dr\,\rho(r)\,e^{-s-srQ}\,.
\end{equation}

As a simple example, let us treat the uniform distribution,
$\rho(r)=1$ when $r\in [0,1]$.  Then Eq.~\eqref{Laplace:int} becomes
\begin{equation}
\label{Laplace:uniform}
s\,e^{s}\widehat{P}(s)= \int_1^\infty dQ\,P(Q)\,\frac{1-e^{-sQ}}{Q}\,.
\end{equation}
Differentiating with respect to $s$ we obtain
\begin{equation}
\label{Pi:uniform}
\frac{d \Pi(s)}{ds} = s^{-1}e^{-s}\,\Pi(s), \qquad \Pi(s) = s\,e^{s}\,\widehat{P}(s)\,,
\end{equation}
whose solution is
\begin{equation}
\label{P:sol}
\widehat{P}(s) = |s|^{-1}\exp[-s-\gamma+\text{Ei}(-s)]\,,
\end{equation}
where $\gamma=0.577\,215\ldots$ is the Euler constant and
$\text{Ei}(\cdot)$ is the exponential integral.

Because this Laplace transform exists for all $s\in \mathbb{R}$,
$P(R)$ decays faster than exponentially in $R$ for $R\to\infty$; this
bound ensures that the Laplace transform \eqref{Laplace} remains
well-defined when $s\to -\infty$.  Using \eqref{P:sol} we find
\begin{eqnarray}
\ln \widehat{P}(-\sigma) = \text{Ei}(\sigma)+\sigma - \ln \sigma -\gamma\,,
\end{eqnarray}
which grows as $\sigma^{-1} e^\sigma$ when $\sigma\gg 1$. This limiting behavior leads to 
\begin{equation}
\label{PR-large}
\ln P(R) \simeq - R\, \ln R 
\end{equation}
for $R\gg 1$.  This is essentially a factorial decay:
$P(R)\propto 1/\Gamma(R)$, where $\Gamma(\cdot)$ is the Euler gamma
function.  This behavior is consistent with the faster than
exponential decay of $P(R)$ observed in simulations
(Fig.~\ref{fig:tn}(b)).  For the small-$R$ behavior, we use the
asymptotic $\widehat{P}(s) \simeq s^{-1}\exp[-s-\gamma]$ as $s\gg 1$
to give $P(1) = e^{-\gamma} = 0.561\,459\ldots$.  This disagrees with
simulations (see Fig.~\ref{fig:tn}(b)), where $\rho(r)$ was chosen
from a uniform distribution in $[a,1]$, with $a=0.9$.  The reason for
this discrepancy is simple: when $\rho(r)$ vanishes for $r<a$, it is
very unlikely to generate a value of $R$ that is close to the minimum
value $R_\text{min}=(1-a)^{-1}$ because it requires each $r_i$ to be
close to $a$.


If the support $[a,b]$ of the distribution $\rho(r)$ is not inside
$[0,1]$, that is, $b>1$, the Kesten variable still has a stationary
distribution if
\begin{equation}
\label{stationary}
\int_a^b dr\,\rho(r)\ln r<0
\end{equation}
Here, the large-$R$ behavior of $P(R)$ is again
algebraic~\cite{Kesten73}, $P(R)\sim R^{-\beta}$, where $\beta$ is the
smallest root of the equation $\int_a^b dr\,\rho(r) r^{\beta-1} = 1$
that also satisfies $\beta>1$. For example, for an arbitrary
distribution with support that is symmetric about $r=1$ (so that it
satisfies $\rho(r)=\rho(2-r)$), the requirement \eqref{stationary} is
always obeyed, so the Kesten variable is stationary.  Here the decay
exponent is universal: $\beta=2$. Thus, already the first moment
$\int dR\, R P(R)$ diverges.

Mitigation strategies are necessarily successful when $\rho(r)$ has
its support inside $[0,1]$.  For distributions $\rho(r)$ defined in
$[a,b]$ with $b>1$, even if the stationarity requirement
\eqref{stationary} is obeyed, the distribution for for the outbreak
size $P(R)$ has an algebraic tail, which implies that a finite
fraction of population contracted the disease.  While the emergence of
these heavy-tailed distributions sparked
interest~\cite{Kesten73,Solomon,Vervaat,Derrida83,JML85,JML88,Nieuwenhuizen}
in Kesten variables, in the context of pandemics, such a feature is to
be avoided.


%

\end{document}